\documentclass[3p,times,procedia]{elsarticle}
\usepackage{nupha_ecrc,url,amssymb}

\volume{00}
\firstpage{1}

\journalname{Nuclear Physics A}
\runauth{D. Oliinychenko}

\jid{nupha}

\jnltitlelogo{Nuclear Physics A}

\begin{document}

\begin{frontmatter}

\dochead{XXVIIIth International Conference on Ultrarelativistic Nucleus-Nucleus Collisions\\ (Quark Matter 2019)}

\title{Overview of light nuclei production in relativistic heavy-ion collisions}


\author{Dmytro Oliinychenko}

\address{Nuclear Science Division, Lawrence Berkeley National Laboratory, 1 Cyclotron Rd., Berkeley, US, 94720}

\begin{abstract}
We briefly overview motivations, some recent results and challenges in studying light nuclei production in relativistic heavy ion collisions. 
\end{abstract}


\end{frontmatter}

\section{Introduction} \label{sec:intro}
Production of light nuclei ($d$, $t$, ${}^3\mathrm{He}$, ${}^3_{\Lambda}\mathrm{H}$, ${}^4\mathrm{He}$, and their antiparticles) in relativistic nucleus-nucleus (AA), proton-nucleus (pA), and pp collisions is studied at least since the early 1960ies \cite{Butler:1961pr,Hagedorn:1960zz}. However, in the last few years this subject has received an increased attention. New precise measurements challenge theoretical approaches to the point that maybe even the mechanism of the light nuclei production has to be reviewed. In these proceedings we attempt to briefly summarize selected experimental results, theoretical models, and how the first challenge the second. For larger overviews see~\cite{Braun-Munzinger:2018hat} (loosely bound objects at LHC) and \cite{Chen:2018tnh} (antinuclei at RHIC and LHC).

\textbf{Light anti-nuclei in cosmic rays:} One reason to investigate light nuclei production stems from the cosmic ray studies. The AMS-02 experiment at the International Space Station \cite{Kounine:2012ega} is measuring fluxes of electrons, positrons, various nuclei, and anti-nuclei in space \cite{Vagelli:2019tqy}. By now no anti-deuterons were reported, but few, yet unpublished, ${}^3\overline{\mathrm{He}}$ and ${}^4\overline{\mathrm{He}}$ events were possibly registered~\cite{Ting:2016,Kounine:2018cla}. It is debated, whether these events are an ordinary background from pp and pA collisions in space, or if they come from exotic sources, such as dark matter annihilations or anti-matter in space. The answer depends crucially on the ordinary background estimates which vary by factor of 4 for anti-deuterons~\cite{Coral:thesis} and by a factor of 10 for anti-helium~\cite{Blum:2017qnn,Poulin:2018wzu}. The largest uncertainty originates from the coalescence model used to predict production of anti-nuclei in pp, pA, and AA collisions. This uncertainty should be reduced by studying (anti-)nuclei production in accelerator experiments.
\begin{figure}
    \centering
    \includegraphics[width=0.6\textwidth]{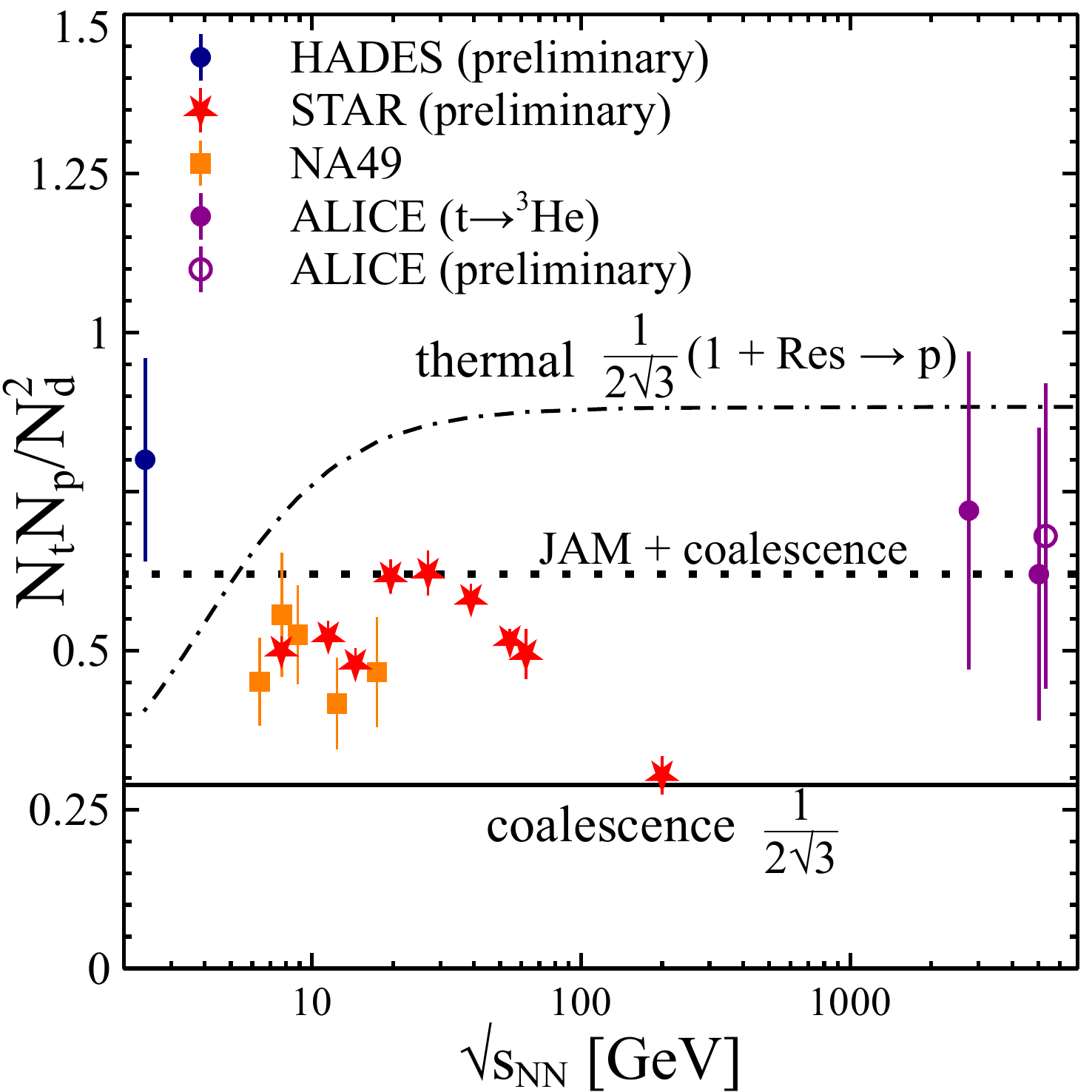}
    \caption{Comparison of the $\frac{N_t N_p}{N_d^2}$ ratio in central PbPb or AuAu collisions at midrapidity between data from HADES (blue circle) \cite{Lorenz_SQM}, NA49 \cite{Anticic:2010mp,Blume:2007kw,Anticic:2016ckv} (squares), STAR \cite{Adam:2019wnb,Zhang:2019wun,Luo:private} (stars), and ALICE \cite{Adam:2015vda} collaborations (purple circles) and models (lines): uniform coalescence (solid) \cite{Sun:2017xrx}, thermal (dash-dotted) \cite{Vovchenko:2019pjl}, and JAM hadronic transport with coalescence afterburner (dotted) \cite{Liu:2019nii}. For the two ALICE data points tritium in the ratio is substituted by ${}^3\mathrm{He}$, which is well-justified at high energies.}
    \label{fig:tpd2_ratio_compilation}
\end{figure}

\textbf{Search for critical point using light nuclei:} Another reason to study light nuclei production is the search for the critical point of strongly interacting matter~\cite{Bzdak:2019pkr}. In the vicinity of the critical point spatial fluctuations of the nucleon density are enhanced. Spatial fluctuations are not measurable directly, but it was recently suggested that the light nuclei production is related to them \cite{Sun:2017xrx,Sun:2018jhg,Shuryak:2018lgd,Shuryak:2019ikv}. This can be illustrated by a simple coalescence model~\cite{Sun:2017xrx}, in which the density is separated into average and fluctuating contributions
 \begin{eqnarray}
   \rho_n(x) = \langle \rho_n \rangle + \delta \rho_n(x) \\
   \rho_p(x) = \langle \rho_p \rangle + \delta \rho_p(x)   
 \end{eqnarray}
and the tritium and deuteron yields are expressed as
 \begin{eqnarray}
   N_d \approx & \frac{3}{2^{1/2}} \left( \frac{2\pi}{mT}\right)^{3/2} \int d^3x \, \rho_p(x) \rho_n(x) &\sim  \left\langle \rho_n \right\rangle N_p (1 + C_{np}) \\
   N_t \approx & \frac{3^{1/2}}{4} \left( \frac{2\pi}{mT}\right)^3 \int d^3x \, \rho_p(x) \rho_n^2(x) &\sim  \left\langle \rho_n \right\rangle^2 N_p (1 + 2 C_{np} + \Delta \rho_n) \,,
 \end{eqnarray}
where
 \begin{eqnarray}
    C_{np} \equiv & \left\langle \delta\rho_n(x) \delta\rho_p(x) \right\rangle / ~(\langle \rho_n\rangle \langle \rho_p \rangle) \\
    \Delta\rho_n \equiv & \left\langle \delta\rho_n(x)^2\right\rangle / \left\langle \rho_n^2 \right\rangle \,.
 \end{eqnarray}
Here $C_{np}$ represents the spatial correlations between neutrons and protons, and $\Delta\rho_n$ corresponds to the spatial fluctuations of neutron density. Therefore the ratio $N_t N_p / N_d^2$ becomes
  \begin{eqnarray} \label{eq:tpd2_ratio}
      \frac{N_t N_p }{ N_d^2} = \frac{1}{2\sqrt{3}} \frac{1+2C_{np} + \Delta \rho_n}{(1+C_{np})^2} 
  \end{eqnarray}
Here the factor $\frac{1}{2\sqrt{3}}$ is related to spin degeneracies and masses, $\frac{g_t g_p}{g_d^2} \left(\frac{3m \cdot m}{(2m)^2}\right)^{3/2} = \frac{1}{2\sqrt{3}}  \approx 0.29$, and the second factor characterizes spatial correlations and fluctuations. Without fluctuations and correlations, this model predicts the $N_t N_p / N_d^2$ ratio to be independent on collision system, energy, centrality, and isospin content. Therefore, the enhancement of the $N_t N_p / N_d^2$ as a function of collision energy may signal enhanced fluctuations, such as the ones generated by the critical point.
In Fig. \ref{fig:tpd2_ratio_compilation} we provide a compilation of the $\frac{N_t N_p}{N_d^2}$ ratio at midrapidity from recent measurements in 0-10\% central AuAu and PbPb collisions. The non-monotonic structure in Fig. \ref{fig:tpd2_ratio_compilation} is evident, but it is impossible to interpret it as a signature of the critical point, because models do not even agree, what the ratio should be without critical point. This disagreement is conceptual: in the coalescence model nuclei are produced at the late stage of collision --- first all resonances decay into nucleons, then nucleons coalesce into nuclei; in the thermal model nuclei are produced earlier, at the hadronic chemical freeze-out --- nuclei are formed first (and the thermal ratio $N_t N_p / N_d^2$ at this moment is the same that for coalescence), then resonances decay into nucleons. Therefore, the ratio $N_t N_p / N_d^2$ is higher by the factor of resonance feed-down into protons, $(1 + Res\to{p})$. In the JAM hadronic transport + coalescence model~\cite{Liu:2019nii} coalescence is performed at $t = 40$ fm/c. This is substantially later than the hadronic chemical freeze-out, but not all resonances decay by this time. Therefore the result lies between the thermal and simple coalescence models. It is remarkable, that the JAM + coalescence result does not depend on the collision energy, even though some correlations (for example due to resonance decays) and thermal fluctuations are present in the model and contribute to Eq.~ (\ref{eq:tpd2_ratio}).

The data in Fig. \ref{fig:tpd2_ratio_compilation} lie between thermal and coalescence models. This indicates, that the formation of nuclei occurs later than the hadronic chemical freeze-out, but earlier than all resonances decay into nucleons. It is certainly possible, that the structure in Fig.~\ref{fig:tpd2_ratio_compilation} is generated not by criticality, but by collision dynamics. For example, in contrast to simple model assumptions, deuterons and tritons may be created at different times on average; these times may be related to local baryon densities and speed of expansion, which change from one collision energy to another. Also, a non-trivial $N_t N_p / N_d^2$ ratio may be generated by nuclear potentials at the late stages of the reaction. Besides, at low energies the feed-down to tritons and deuterons from the numerous excited states of ${}^4\mathrm{He}$ is important~\cite{Shuryak:2019ikv,Vovchenko:2019},
but it is not taken into account in any of the available models. Furthermore, even if spatial fluctuations of the neutron density are generated by the criticality, it is not clear if they are preserved until the moment, when the light nuclei are created. The above concerns can be clarified within a dynamical approach treating light nuclei explicitly~\cite{Oliinychenko:2018ugs,Oliinychenko:2018odl,Oh:2009gx} or supplied by coalescence~\cite{Liu:2019nii,Zhu:2015voa,Dong:2018cye,Sombun:2018yqh}.

\section{Inventory of present models of light nuclei production} It is clear from Fig.~\ref{fig:tpd2_ratio_compilation}, that the current models of light nuclei production need to be improved to match the precision of the new data. To understand, what can be improved, let us make a short overview of the models: various types of coalescence, thermal + blast wave models, and dynamical models. All coalescence models assume that light nuclei are formed at the late stage of collision from nucleons that reside close enough in the phase space. Here we classify coalescence models into 3 groups: simple analytical coalescence (no nucleus wavefunction is included) \cite{Gutbrod:1988gt,Csernai:1986qf}, advanced analytical coalescence (wavefunction is taken into account) \cite{Sato:1981ez,Scheibl:1998tk,Mrowczynski:2016xqm,Sun:2017ooe}, and dynamical model + coalescence \cite{Zhu:2015voa,Dong:2018cye,Liu:2019nii,Ivanov:2017nae,Sombun:2018yqh}. In thermal + blast wave models spectra are computed in a blast-wave model at hadronic kinetic freeze-out, while the yields to normalize them are taken at hadronic chemical freeze-out. Purely dynamical models are transport approaches, where light nuclei are either forming from nucleons via potentials, or are treated as single degrees of freedom.

\textbf{Simple analytical coalescence:} In a simple coalescence model the phase space density of the nucleus with mass $A$ is proportional to the product of nucleon phase space densities:
\begin{eqnarray}
  E_{A} \frac{d N_{A}}{d^{3} P_{A}}=B_{A}\left(E_{\mathrm{p}} \frac{d N_{\mathrm{p}}}{d^{3} P_{\mathrm{p}}}\right)^{Z}\left.\left(E_{\mathrm{n}} \frac{d N_{\mathrm{n}}}{d^{3} P_{\mathrm{n}}}\right)^{N}\right|_{P_{\mathrm{p}}=P_{\mathrm{n}}=P_{A} / A}
\end{eqnarray}
It turns out from measurements that the coalescence parameter $B_A$ depends on the nucleus mass number $A$, collision system, centrality, energy, and transverse momentum. These dependencies are qualitatively reproduced, if one observes that by dimension $B_A  m \sim V^{-(A-1)}$, where $V$ has a dimension of volume. Then, using the measured dependencies of HBT-volume $V_{HBT}$, and taking $V_{HBT}$ as a proxy for the volume $V$, simple coalescence predicts that
\begin{itemize}
    \item $B_3/B_2^2$ does not depend on collision system size, centrality and $p_T$, in fact it is a differential analogue of $N_{t} N_{p} / N_{d}^2$ ratio
    \item $B_A(p_T)$ grows with $p_T$ in AA, $B_A(p_T) \approx const$ in pp
    \item $B_A$ decreases with larger event multiplicity
\end{itemize}
Simple coalescence also predicts approximate flow scaling $v_2^A(p_T) = A v_2(p_T/A)$. These predictions are indeed fulfilled qualitatively. Measurements by the ALICE collaboration clearly show that $B_2$ and $B_3$ stay almost constant as a function of $p_T$ in pp and peripheral PbPb collisions, but grow in more central collisions \cite{Acharya:2019rgc,Adam:2015vda}. The same trend is measured at lower energies by STAR and NA49~\cite{Anticic:2016ckv,Adam:2019wnb}. The dependence of $B_2$ on charged particle multiplicity in pp, pA, and AA is also measured by ALICE~\cite{Puccio:2019oyd}: the larger the multiplicity, the smaller $B_2$. The coalescence scaling of $v_2$ is fulfilled only approximately, mainly at small $p_T$~\cite{Haque:2017qhe}. Overall, simple coalescence provides correct qualitative explanations, but for more precise quantitative predictions it has to be refined.

\textbf{Advanced analytical coalescence:} We call an analytical coalescence approach advanced, if it takes into account the wavefunction of the produced nucleus. A well-known example of such approach is~\cite{Scheibl:1998tk}, which was recently rederived with relaxed assumptions~\cite{Blum:2019suo}. The expression for the deuteron yield in advanced analytical coalescence involves the Wigner-function of deuteron $\mathcal{D}_d(\vec{q}, \vec{r})$:
\begin{eqnarray}
  \frac{dN_d}{d^3P_d} = G_d \int d^3R \int \frac{d^3q}{(2\pi)^3} \int d^3r \, \mathcal{D}_d(\vec{q},\vec{r}) \times \mathit{f}\left(\frac{\vec{P}_d}{2} + \vec{q}, \vec{R} + \frac{\vec{r}}{2}\right) \mathit{f}\left(\frac{\vec{P}_d}{2} - \vec{q}, \vec{R} - \frac{\vec{r}}{2}\right) \,,
\end{eqnarray}
where $\vec{r}$ is a separation between nucleons in coordinate space and $\vec{q}$ is a separation in momentum space. Neglecting $\vec{q}$ compared to the deuteron momentum $\vec{P_d}$ and using the deuteron wavefunction $\phi_d(r)$ satisfying $(2\pi)^3 |\phi_d(\vec{r})|^2 = \int \mathcal{D}_d(\vec{q},\vec{r}) d^3q$ one can obtain the connection between the deuteron production and nucleon-nucleon correlations $C_2(\vec{p},\vec{q})$~\cite{Blum:2019suo}:
\begin{eqnarray}
  B_2(\vec{p}) \approx \frac{3}{2m} \int d^3q \, \mathcal{D}(\vec{q}) \, C_2(\vec{p},\vec{q}) \,,
\end{eqnarray}
where $\mathcal{D}(\vec{q}) = \int |\phi_d(\vec{r})|^2 e^{i \vec{k}\vec{r}}$. To proceed further analytically one assumes Gaussian wavefunction and Gaussian source model
\begin{eqnarray}
  \mathcal{D}(\vec{k}) = e^{-k^2 d^2 /4} \\
  C_2(\vec{p},\vec{q}) = e^{-R_{\perp}^2 \vec{q}_{\perp}^2 - R_{\parallel}^2 q_{\parallel}^2}
\end{eqnarray}
Under these assumptions one obtains
\begin{eqnarray}
  B_2(\vec{p}) = \frac{3\pi^{3/2}}{2m  \left(R_{\perp}^2(p) + d^2/4 \right) \left( R_{\parallel}^2(p) + d^2/4\right)^{1/2}}
\end{eqnarray}

If the fireball is large enough compared to deuteron size $d$, $d \ll R_{\perp}$, $d \ll R_{parallel}$, then one retrieves all the results of a simple coalescence. However, here one more feature is present: in a small system, such as pp or pA, the deuteron size plays a role. One can extend these results to larger nuclei, assuming that many-nucleon correlations can be represented as a sum of the pairwise Gaussian correlations. Also one has to assume an $(A - 1)$-dimensional symmetric Gaussian form for the cluster's relative coordinate wave function. Then
\begin{eqnarray}
  B_A = \frac{(2 J_A + 1)}{2^A \sqrt{A}} \left[\frac{(2\pi)^{3/2}}{m  \left(R_{\perp}^2(p) + d^2/4 \right) \left( R_{\parallel}^2(p) + d^2/4\right)^{1/2}}\right]^{A-1} \,,
\end{eqnarray}
where $2 J_A + 1$ is a spin degeneracy. Altogether this implies that nuclei with the large wavefunctions are suppressed, especially so in smaller systems. Based on this it was recently suggested \cite{Bellini:2018epz} to use $^{3}_{\Lambda}\mathrm{H}$ to test advanced coalescence. The $\Lambda$ is bound to $pn$ pair by only 130 keV in $^{3}_{\Lambda}\mathrm{H}$. Therefore, the wavefunction of $^{3}_{\Lambda}\mathrm{H}$ is extended to more than 10 fm, which is comparable to a fireball size in AA collisions and exceeds the fireball size in pA and pp. Therefore, advanced analytical coalescence predicts considerable suppression of $^{3}_{\Lambda}\mathrm{H}$~\cite{Bellini:2018epz}. In Pb+Pb collisions at 2.76 TeV this suppression is not observed. However, it turns out that a coalescence model with a more careful consideration of an $^{3}_{\Lambda}\mathrm{H}$ wavefunction instead of Gaussian ansatz is not in tension with experiment~\cite{Zhang:2018euf}. To sum up, the question whether the spatial extent of the wavefunction matters is still open. Measurements of $^{3}_{\Lambda}\mathrm{H}$ by STAR would provide new input to it and challenge the existing models.

\textbf{Dynamical approach + coalescence:} In analytical coalescence a number of assumptions was made ($|\vec{q}| \ll |\vec{P}_d|$, Gaussian source, Gaussian wavefunction) to make the model approachable analytically. In a dynamical model + coalescence approaches these assumptions are not necessary. Light nuclei are constructed directly from the nucleons originating from a Monte-Carlo transport (or hydrodynamics + transport) simulation, if nucleons are close enough in the phase space \cite{Zhu:2015voa,Dong:2018cye,Liu:2019nii,Sombun:2018yqh}. This takes into account a realistic space-time distribution of the nucleons. To give an example of such model, consider the algorithm from \cite{Sombun:2018yqh}: one loops over all $pn$ pairs from the UrQMD transport approach, each single pair is traced back in time to the latest of interaction times (when both nucleons are frozen out), boosted to its center of mass frame, and considered bound if the distance between nucleons does not exceed 0.28 GeV in momentum and 3.5 fm in coordinate space. Then spin and isospin factors are taken into account. This scheme contains two adjustable parameters, and it allows to describe deuteron production in AA from HADES to LHC, and reproduces multiplicity dependence of deuteron production at LHC from pp to AA~\cite{Sombun:2018yqh}.

The downside, however, is that the way of constructing light nuclei is algorithmic and varies from model to model.
Furthermore, in some cases (for example \cite{Liu:2019nii}) coalescence is performed at fixed time, which makes results dependent on this time.

\textbf{Thermal + blast wave model:} The thermal model assumes a hadron resonance gas in a global chemical equilibrium until a sharp chemical freeze-out --- the moment, when interaction changing hadron yields cease. Therefore, final hadron yields $N_i$ of hadron $i$ are a sum of thermal yields and feed-down from resonance decays:
\begin{eqnarray}
  N_i^{thermal} = \frac{g_i V e^{\mu_i/T_{ch}}}{2\pi^2 \hbar^3} m_i^2 T_{ch} K_2\left(m_i/T_{ch}\right)\\
  N_i = N_i^{thermal} + \sum_R N_R^{thermal} br(R \to i) \,,
\end{eqnarray}
where $m_i$ is hadron mass, $g_i$ is it's degeneracy, $V$ is the volume of the system, $br(R\to i)$ is an average amount of hadron $i$ from decay of a resonance $R$, $T_{ch}$ is the temperature of the chemical freeze-out, $\mu_i = \mu_B B_i + \mu_S S_i + \mu_{I3} I_{3i}$ is a chemical potential of hadron $i$, with $ \mu_B$, $\mu_S$, and $\mu_{I3}$ being baryon, strangeness and isospin chemical potentials. The latter two are adjusted to satisfy net strangeness neutrality and isospin to baryon ratio $\approx 0.4$ as in the initial nuclei. To describe not only the yields, but also the spectra, the thermal model is supplied by the collective flow resulting in \cite{Schnedermann:1993ws}
\begin{eqnarray} \label{eq:blastwave}
  \frac{dN}{p_T dp_T} \sim \int_{0}^{R} K_1\left(\frac{m_T}{T_{kin}} \cosh \rho \right) I_0\left(\frac{p_T}{T_{kin}} \sinh \rho \right)r dr \\
  \tanh \rho = \beta_s \left(\frac{r}{R} \right)^n \,,
\end{eqnarray}
where $R$ is the source radius and $\beta_s$ is the collective radial expansion velocity. The parameters $n$, $R$, $\beta_s$, $T_{ch}$, $T_{kin}$, $\mu_B$, $V$ are obtained from a combined fit of hadrons and light nuclei.

The thermal model provides an excellent description of hadronic yields from AGS ($E_{kin} = $ 2-10 AGeV) to ALICE energies ($\sqrt{s_{NN}} =$ 2.76 and 5.02 TeV)~\cite{Andronic:2005yp,Andronic:2017pug}. The light nuclei yields measured by ALICE --- $d$, $^3\mathrm{He}$, $^3_{\Lambda}\mathrm{H}$, and even $^4\mathrm{He}$ --- are described by the thermal model as well~\cite{Adam:2015vda,Adam:2015yta,Acharya:2017bso}, and the chemical freeze-out temperature is the same for nuclei and hadrons. Deuteron midrapidity yields at the STAR beam energy scan energies (7.7 -- 200 GeV) are described fairly well by the thermal model, while the yield of $t$ seems to be overestimated by roughly a factor of two~\cite{Zhang:2019wun}. The measured nuclei spectra in Pb+Pb collisions at ALICE are in agreement with the blast-wave model~\cite{Adam:2015vda}, but in p+Pb it significantly overestimates deuteron mean transverse momentum $\langle p_T \rangle$~\cite{Acharya:2019rys}. Qualitatively, the blast wave model reproduces the growth of $B_2(p_T)$.

A good description of the light nuclei yields in AA collisions by the thermal model generated debates, whether light nuclei may be produced at the hadronization directly from quarks and gluons. Such production mechanism is in conceptual conflict with the coalescence approach, which assumes that nuclei are produced at the late stage of the fireball expansion. However, distinguishing these mechanisms unambiguously appears to be an unsolved problem, despite several attempts to approach it~\cite{Bazak:2018hgl,Mrowczynski:2019yrr,Bellini:2018epz}. In fact, it is possible that both mechanisms are at play: light nuclei can be produced at hadronisation, part of them being destroyed during the hadronic evolution, while the new ones being created via coalescence at the later stages of hadronic evolution. Simulations suggest that at LHC not more than 20\% of deuterons are from hadronisation~\cite{Oliinychenko:2018ugs}.

The thermal model relies on the fact that inelastic reactions changing hadronic yields (for example $NN \to N\Delta \to NN\pi$) have smaller cross sections than those of quasi-elastic reactions that do not change hadronic yields (for example $\pi N \to \Delta \to \pi N$). As a consequence, the inelastic reactions cease first -- a chemical freeze-out occurs. The yields are not changing after this. The quasi-elastic reactions cease later -- this is the kinetic freeze-out, when the momentum spectra are frozen. When the same concept is applied to the light nuclei, the initial assumption is not true anymore: inelastic cross sections are larger for them than elastic. In other words, hadrons are more likely to disintegrate the light nuclei, rather than scatter off elastically. This poses a ``snowballs in hell'' question: how do light nuclei survive from chemical to kinetic freeze-out. Dynamical simulations~\cite{Oliinychenko:2018ugs}, as well as analytical models~\cite{Xu:2018jff,Vovchenko:2019aoz} suggest that they do not survive, but are disintegrated and re-created at similar rates, and therefore remain in relative equilibrium with nucleons.

\textbf{Comparing blast wave model and coalescence model:} A question is repeatedly posed, how one can discriminate the thermal + blast wave model from the coalescence model~\cite{Bazak:2018hgl,Mrowczynski:2019yrr,Bellini:2018epz}. Let us demonstrate, why this is a difficult problem and suggest, how it can be resolved. The blast-wave model spectra (Eq. \ref{eq:blastwave}) are derived by locally boosting a thermal Boltzmann distribution and integrating the result over the particular boost-invariant hypersurface of kinetic freeze-out. Instead, here we wish to compare the blast wave model and coalescence on a \textit{generic} hypersurface of chemical freeze-out $\Sigma$:

\begin{eqnarray}
  \frac{dN_d}{d^3p} = \frac{g_d}{(2\pi)^3} \int_{\Sigma} \frac{p^{\mu} d\sigma_{\mu}}{p^0} \exp \left(   \frac{\mu_d - p^{\mu}u_{\mu}}{T_{kin}} \right) \,.
\end{eqnarray}

Here one assumes a Boltzmann distribution rather than Fermi, which is well-justified because $m_{N} \gg T_{kin}$. This should be compared to analogous expression in the coalescence model, where the hypersurface of coalescence is not yet specified (Eqs. 3.19--3.20 of~\cite{{Scheibl:1998tk}}):

\begin{eqnarray}
  \frac{dN_d}{d^3p}  = \frac{g_d}{(2\pi)^3} \int_{\Sigma} \frac{p^{\mu} d\sigma_{\mu} (R_d)}{p^0} \mathit{f}_p \left(R_d, \frac{p}{2}\right) \mathit{f}_n \left(R_d, \frac{p}{2}\right) C_d(R_d, p) \,,
\end{eqnarray}

where $C_d$ is a quantum mechanical correction factor, which is equal to unity if the nucleon density is uniform on the scale of the deuteron size. Typical values of $C_d$ in heavy ion collisions are 0.8--0.9. Assuming that
\begin{eqnarray}
  \mathit{f}_{p,n} \left(R_d, \frac{p}{2}\right) = \exp \left(   \frac{\mu_{p,n} - p^{\mu}u_{\mu}/2}{T_{kin}} \right) \,,
\end{eqnarray}
which is the assumption of the blast-wave model, and neglecting the deuteron binding energy compared to its mass, one obtains that
\begin{eqnarray}
  \mathit{f}_p \left(R_d, \frac{p}{2}\right) \mathit{f}_n \left(R_d, \frac{p}{2}\right) = \exp \left(   \frac{\mu_d - p^{\mu}u_{\mu}}{T_{kin}} \right) \,.
\end{eqnarray}

This means that the blast-wave model and coalescence model essentially differ only by the quantum correction factor $C_d$. If the nucleon density remains approximately unchanged on the scale of the deuteron size, then $C_d \approx 1$ and the models are indistinguishable. Similar results are obtained for heavier nuclei. Given the same hypersurface, the coalescence model is nothing else but an advanced version of the blast wave + thermal model, which additionally accounts for density fluctuations and the wavefunction of light nuclei. If one can experimentally demonstrate the connection between density fluctuations and light nuclei production, which is absent in the thermal model, this will be the ultimate success of the coalescence model.

\section{Summary}

Studying light nuclei production in relativistic ion collisions helps to understand the anti-nuclei abundance in space and can potentially give clues about lumps of antimatter in space. It also may allow to pinpoint the critical point of the strongly-interacting matter. The latter can be done by studying the ratio $\frac{t p}{d^2}$ as a function of collision energy. It has been recently measured with an excellent precision and, as shown in Fig. \ref{fig:tpd2_ratio_compilation}, it exhibits a non-trivial structure, which cannot be explained by any of the current models.

Models do not agree whether nuclei are formed earlier, before all resonances decay into nucleons, or later, when all resonances have already decayed into nucleons. These scenarios can be distinguished by a more precise measurement of the $\frac{N_t N_p}{N_d^2}$ ratio. Recent STAR data favors an intermediate case, where only a fraction of the resonances decays by the moment nuclei are formed.

To improve current models of light nuclei productions, the crucial step is to better understand the role of nuclei wavefunction and spatial fluctuations of the nuclear density. Both play a role in coalescence models, but do not matter in the thermal model. To address this, it is helpful to consider the production of $^3_{\Lambda}\mathrm{H}$ in pp, pA, and AA both experimentally and theoretically, because $^3_{\Lambda}\mathrm{H}$ spatial extent is particularly large.

The role of fluctuations is to be further studied in the dynamical + coalescence models and in purely dynamical models. The latter would profit from measured or precisely computed cross sections of light nuclei reactions with hadrons, such as $\pi+d$, $\pi + t$, $\pi + {}^3_{\Lambda}\mathrm{H}$, $N + d$, $N + t$, $N+ {}^3_{\Lambda}\mathrm{H}$, etc.

\paragraph{Acknowledgements}

We would like to thank X. Luo for sharing the STAR data shown in Fig. 1, and to the hepdata project (hepdata.net), that allowed to seamlessly obtain NA49 and ALICE data shown in the same Figure. D.O. thanks B. D\"onigus, V. Vovchenko, E. Shuryak, H. Elfner, L.-G. Pang, and V. Koch for fruitful discussions. This work was supported by the U.S. Department of Energy,
Office of Science, Office of Nuclear Physics, under contract number 
DE-AC02-05CH11231 and received support within the framework of the
Beam Energy Scan Theory (BEST) Topical Collaboration.

\bibliographystyle{elsarticle-num}
\bibliography{inspire,noninspire}

\end{document}